\documentstyle[12pt]{article}\begin{document}
\title{Towards an unified description of total and diffractive structure functions at
HERA in the QCD dipole picture}
 \author{
A. Bialas\thanks{%
Institute of Physics, Jagellonian University, Reymonta 4, 30-059 Cracow,
Poland}, 
R. Peschanski\thanks{%
CEA, Service de Physique
Th\' eorique, CE-Saclay, F-91191 Gif-sur-Yvette Cedex, France}
 and Ch. Royon\thanks{%
CEA, DAPNIA, Service de Physique des Particules, 
CE-Saclay, F-91191 Gif-sur-Yvette, France}}
\maketitle
\begin{abstract}
It is argued that the QCD dipole picture allows to build an unified 
theoretical description -based on BFKL dynamics- of the total and 
diffractive nucleon structure functions. This description is in qualitative agreement with the present 
collection of data obtained by the H1 collaboration.  
More precise theoretical estimates, in particular the determination of the  normalizations and  proton transverse momentum behaviour of the diffractive components, are shown to be required in order to reach definite conclusions.
\end{abstract}
\bigskip
\section{Motivation}

Considering the phenomenological discussion on the proton structure functions measured  by deep-inelastic scattering of
electrons and positrons at HERA, it is striking to realize that the proposed
models, on one side for the total quark structure function $F_{2}\left(
x,Q^{2}\right) $ \cite{H1} and on the other side for its diffractive component $%
F_{2}^{D\left( 3\right) }\left( x,M^{2},Q^{2}\right) $ \cite{F2DH1} are 
in general distinct.
Indeed, the models  \cite{review} aiming at the description of $F_{2}\left(
x,Q^{2}\right) $ use a QCD-inspired ``hard Pomeron''
parametrisation related either to a DGLAP \cite{DGLAP}
evolution with extrapolation at small-$x$ \cite{GRV} or to
BFKL \cite{BFKL} dynamics.
On the other hand, most of the models proposed for the diffractive component of the
quark structure function  rely on a ``soft
Pomeron'' picture of diffraction, assuming a point-like structure of the Pomeron considered as a compound particle \cite{Ingelman}.

It is known since some time, however, that at high energies the elastic scattering and diffraction dissociation of hadrons are closely related \cite{good}, being both a reflection of the same phenomenon, namely {\it absorption} of the incident particle wave in the target.  It seems therefore interesting to verify if the same applies also to the incident virtual photons.

In the present paper we investigate this question in the framework of the QCD   dipole picture
\cite{Mueller,Niko}. This picture turned out already to be successful in the description of the 
total virtual photon-nucleon cross-section (i.e. of total nucleon structure
function $F_2$ \cite{ourpap}). The purpose of the present paper is to verify if the so-called rapidity gap events \cite{review} discovered recently at HERA can also be described along these lines.
Diffractive dissociation of the virtual photons in the framework of the QCD dipole picture was recently discussed in \cite{inelcomp,elcomp}. It was argued that the diffractive cross-section consists of two components:

-(component I) the {\it inelastic} component when the gluon cascade which evolved from the incident virtual photon interacts inelastically with the target, see Fig.1a. This component (corresponding to the 3-Pomeron interaction in the Regge terminology) contributes mainly to the region
 of very large mass $M$ of the diffractively excited system: $\beta <\!< 1$, where, as usual, $\beta=Q^2/(Q^2+M^2)$.

-(component II) the {\it quasi-elastic} component when the $q\bar{q}$ pair emerging from the
virtual photon scatters elastically from the target, see Fig.1b. This component contributes
to the region of smaller masses $\beta \geq .2$ .

The model calculations of Refs. \cite{inelcomp,elcomp} provided the formulae for differential cross-section $d\sigma/dM^2$ of both components I and II.
Unfortunately, for technical reasons, some rather drastic assumptions had to be made:

(a) The calculations were performed in the limit of  large impact 
parameters. The integrated cross-section was then estimated by  integration only up to a certain cut-off $b_{min}$. This procedure leads to a serious underestimate of the cross-section \cite{bia}.

(b) The target nucleon was treated as a collection of several QCD dipoles 
all of the same size and sitting at one point. This assumption neglects the effects of nucleon form-factor and thus leads to an overestimation of the cross-sections.

We are thus led to the conclusion that there are at present no reliable predictions for the {\it absolute normalization} of the diffractive cross-sections
of the virtual photons. In this situation, in order to compare the predictions with the data we  decided to  treat the normalization
constants in the two components as arbitrary parameters  and restrict ourselves to the comparison of the observed dependence on kinematic variables to that 
predicted by the formulae of  Refs. \cite{inelcomp,elcomp} (c.f. also \cite{ge1}). Our work should thus be treated as an exploratory search which is a guide for further investigation and should be repeated once  more reliable  
calculations are available. Within these {\it caveats} our investigation leads to the conclusion that the
data on rapidity gap events published recently by H1 collaboration \cite{F2DH1}
are reasonably well described by the QCD dipole picture and thus the Good-Walker idea seems consistent with these data.

The plan of our investigations is as follows. In the next section we remind 
briefly the QCD dipole picture results for the total photon-nucleon 
cross-section, introduce the necessary notation and perform a fit for the total structure function. In Section 3 we summarize the
 formulae  for diffractive cross-section of the components (I) and (II). These 
results are compared to the data \cite{F2DH1} in Section 4. 
 Finally Section 5 contains our conclusions, as well as an outlook for  further work.

\section{Proton structure functions}

In the  QCD-dipole picture   of high-energy
scattering of two initial small-size $\left( r,\bar{r}\right) $ {\it onia} (massive $q\bar{q}$ states), the
total cross-section  at fixed impact parameter $\sigma(b)$ can be obtained from the
all-order  QCD resummation of  the elementary dipole-dipole 
cross-sections $\sigma\left( \rho,\bar{\rho}\right) $ where dipole states of transverse diameter $\rho$ (resp. $%
\bar{\rho}$) appear in the wave-function of the initial states of transverse diameter  $r$ (resp.$%
\bar{r}$) at the ``time'' of interaction. This ``time'' variable is represented by
a rapidity variable $\ln c/\xi ,$ where $c$ is a phenomenological constant  
\cite{Bialas} and $%
\xi $ is the Bjorken variable labelling the softer end of the produced dipole
.
One writes
\begin{eqnarray}
\sigma_{tot}= \int d^2b\ \sigma(b) {=}\int \frac{d\rho}{\rho}%
 n_{1}\left(r;\rho,\xi \right) \int \frac{d\bar{\rho}}{\bar{\rho}}
n_{1}\left(\bar{r},\bar{\rho},\bar{\xi}\right) \ \sigma \left( \rho,\bar{\rho}\right) , 
\label{1}
\end{eqnarray}
\noindent where the partition   of the total  ``time'' $\ln c/x = \ln c/\xi \bar \xi $ between the target and projectile  is arbitrary, provided $\xi \bar\xi =x$ .  
$n_{1}\left( r,\rho,\xi \right)$  
 is the   multiplicity of dipoles of
size $\rho,$ integrated over the transverse distance from the center of the onium, generated from an initial
dipole of size $r$ after a ``time'' $\ln c/\xi .$ It is given by
\begin{equation}
n_1(r,\rho,\xi)= \int \frac{d\gamma}{2\pi i}\ \left(\frac{r}{\rho}\right)^{\gamma}
\ e^{\Delta(\gamma)\ln(1/\xi)}
\label{2a} 
\end{equation}
where
\begin{equation}
\Delta(\gamma) = \frac{\alpha N}{\pi}(2\psi(1) - \psi(1-\gamma/2)-\psi(\gamma/2))  
\label{2aa}
\end{equation}
is the eigenvalue of the BFKL kernel \cite{BFKL}, and $N=3$ is the number of colours.

The elementary dipole-dipole cross-sections are  obtained from the
gluon-exchange graphs and give \cite{Mueller}
\begin{eqnarray}
\sigma\left( \rho,\bar{\rho}\right) =8\pi \alpha ^{2}\int \frac{d\ell }{\ell ^{3}}%
\left[ 1-{J}_{0}\left( \ell \rho\right) \right] \left[ 1-{J%
}_{0}\left( \ell \bar \rho\right) \right]i\nonumber \\ =2\pi \alpha^2 \rho_<^2\left[1+\ln(\rho_>/\rho_<)\right]  .  \label{4}
\end{eqnarray}

\noindent Inserting formulae (\ref{2a}) and (4) in the cross-section formula (1),
 one finds
\begin{equation}
\sigma(r,{\bar r},\xi)= 2\pi \alpha^2 r {\bar r} \int \frac{d\gamma}{2\pi i}\left(\frac{r}{{\bar r}}\right)^{\gamma-1} \frac{4}{\gamma^2(2-\gamma)^2}
 \ e^{\Delta(\gamma)\ln(1/\xi)}
\label{5a} 
\end{equation}
In order to obtain the virtual photon-proton cross-section from (\ref{5a}) one has to integrate over the initial distributions of dipoles inside the photon
and the proton. Since we know neither the number nor the distribution of the
dipoles in the proton, we simply define
\begin{equation}
\int d^2{\bar r}\  ({\bar r})^{2-\gamma}\ \Phi({\bar r}) \equiv \ n_{eff}(\gamma) \ [r_0(\gamma)]^{2-\gamma}
\label{5b}
\end{equation}
where $n_{eff}$ has the meaning of the average number of primary dipoles in
the proton and  $r_0$ is their average transverse diameter.

The distributions of the primary dipoles in the virtual photons are known \cite{Bjorken,Niko} and 
thus the corresponding integrals can be performed with the result
\begin{eqnarray}
\sigma_{T,L}(x,Q^2)= \frac{4 N \alpha_{em} e_f^2}{\pi} \alpha^2 n_{eff}  \int \frac{d\gamma}{2\pi i} r_0^2\ \left(\frac{2}{Qr_0}\right)^{\gamma} \ e^{\Delta(\gamma)\ln(1/x)}
\nonumber \\ \times  \                              \frac{4}{\gamma^2(2-\gamma)^2}  
\frac{\Gamma^2(2-\gamma/2)\Gamma^4(1+\gamma/2)}{\Gamma(4-\gamma)\Gamma(2+\gamma)}\ H_{T,L}(\gamma)   \label{5c}
\end{eqnarray}
where
\begin{equation}
H_T(\gamma)= \frac{(2-\gamma/2)(1+\gamma/2)}{\gamma(1-\gamma/2)},\;\;H_L(\gamma)=1 . 
\label{19}
\end{equation}
$H_{T,L}$ refers to 
transverse and longitudinal photons, respectively.
$e_{f}^{2}$ is the total charge of the quarks whose flavour contributes to
the reaction, and $n_{eff}=n_{eff}(1),\; r_0=r_0(1),$ see formula (\ref{5b}).

The path integral in (\ref{5c}) can be evaluated by the saddle point method
(giving good approximation as $x\rightarrow 0$). The result is
\begin{eqnarray}
F_{T,L} = \frac{Q^2}{4\pi^2\alpha_{em}} \sigma_{T,L}=H_{T,L}(1)\ \frac{ \pi N \alpha^2  e_f^2}{32} 
\ n_{eff}
\left( 
\frac{x}{c}\right) ^{-\Delta _{ P}}\frac{r_0Q}{2} \nonumber \\ \times 
\left( \frac{2a\left(x\right) }{\pi}\right) ^{1/2}\exp \left\{ -\frac{a\left( x\right)}{2} \ln ^2 \left( \frac{r_0 Q}{2}
\right) \right\} .  
\label{5x}
\end{eqnarray}
where $H_L(1)=1,$ $H_T(1)=9/2$ and
\begin{eqnarray}
a(\xi )=\left[7\alpha N\zeta(3)\ln (c/\xi )/\pi \right] ^{-1}, \;\; 
\Delta _{ P} \equiv \Delta(1) =4\ln 2\;\alpha N/\pi  \label{3}
\end{eqnarray}
\noindent are the well-known coefficients appearing in the solution of BFKL dynamics for the Pomeron \cite{BFKL}.

Formula (\ref{5x}) gives the prediction for the  nucleon structure functions in terms of four  parameters:
the strong coupling constant $\alpha$, the average number of primary dipoles in the proton $n_{eff}$, their average radius $r_0,$ and the constant $c$ fixing the rapidity scale of the problem.
 It  coincides
with the one used in the published fit \cite{ourpap}
apart the new parameter $c$ which sets the
rapidity scale of the process, and is unavoidable in the leading log approximation of QCD. This justifies a new fit of $F_2$ using formula
(\ref{5x}) which we have performed
 assuming $\Delta_P=.282$ (as in \cite{ourpap})   and leaving free the three other parameters. The result is
\begin{equation}
\Delta_P = .282,\;\; c=1.75,\;\; Q_0= \frac{2}{r_0}= .622 {\rm GeV},\;\; n_{eff}= 3.8/e_f^2.                        \label{6}
\end{equation}

The fit (displayed in Fig.2) is using the published data from the H1 experiment
\cite{H1} .
We have only considered the points with $Q^{2} \leq 150 {\rm {\rm GeV}}^{2}$ to remain
in a reasonable domain of validity of the 
QCD dipole model. Changing this value does not appreciably change the quality and parameters of the fit. The $\chi^{2}$ is 88.7
for 130 points. Although not included in the fit, the data points at high 
$Q^{2} > 150 {\rm GeV}^{2}$ and $x < 5. 10^{-1}$ are well described, while at higher $x,$ an expected  contribution of valence quarks is  needed.

Commenting on the parameters, let us note that  the effective coupling constant extracted  from $\Delta_{P}$ is $\alpha =0.11$, close
to $\alpha (M_{Z})$ used in the H1 QCD fit. It is an acceptable value for the small  coupling constant required by the  BFKL framework\footnote  {The running
of the coupling constant and other next leading log corrections are  not taken into account in the present BFKL scheme.  This could explain the rather low value of the effective $\Delta _{ P}$ 
which is expected to be decreased by the next leading contributions \cite {cia}.}.  The value of $Q_{0}$ corresponds
to a transverse radius of $0.4 fm$ which is in the correct range for a proton non-perturbative characteristic 
scale. The value of $n_{eff}$ determines the number  of primordial dipoles in the proton to be about $6$ (if three flavours contribute to the process) which also does not seem unreasonable. 
The parameter $c$ sets the ``time'' scale for the formation of the interacting dipoles. It defines the effective total rapidity interval which is $\ln (1/x) + \ln c,$ the constant being not predictable (but of order 1) at the leading logarithmic
approximation. 
\par
The obtained fit for $F_2$ is very similar  than the previously published one in Ref. \cite{ourpap},  even with  a better $\chi^2$. 
In the same spirit, relation (11) provides a parameter-free
prediction for the gluon density (not shown in the figures) 
which is, as the previous one \cite{ourpap}, in good
agreement with the results obtained by the H1 QCD fits based on a NLO DGLAP
evolution equation \cite{H1}. Using the factorization properties of formula (11) and noting
\cite{ourpap} that the $F_{L}$ structure function  is given by a
similar formula with $h_{T} + h_{L}$  replaced by $h_{L},$ one obtains a parameter-free prediction for $F_{L}$ (see Fig.3). Note that we obtain a prediction  in agreement with the  (indirect)
experimental determination for  $%
F_{L}$ \cite{FLH1}, but somewhat lower than the center values. Thus, it would be interesting to obtain a more precise measurement of $F_{L}$ to
test the different predictions on the $Q^2$-evolution 
as already mentionned in Ref. \cite{ourpap}.

\section{Diffractive structure functions}

The diffractive structure functions are related to the corresponding diffractive $\gamma^*$-nucleon cross-sections by the relation
\begin{eqnarray}
F_{T,L}^{D\left( 3\right) }\left( Q^{2},x_{ P},\beta \right) \equiv \frac{%
Q^{2}}{4\pi ^{2}\alpha _{{\rm e.m.}}}x_{ P}^{-1}\int d^{2}b\ \frac{\beta
\;d\sigma _{T,L}}{d\beta \;d^{2}b},  \label{15}
\end{eqnarray}
where $x_P=x/\beta$.

As already explained in the first section, in the QCD dipole model the diffractive structure functions are given by two components: inelastic and quasi-elastic. They were discussed
in \cite{inelcomp,elcomp} where the formulae for $\gamma^*$-dipole diffractive cross-sections were derived
and used to construct the corresponding structure functions following the formula.
These results are summarized below.
\vspace{0.5cm}

(I) Inelastic component.

\begin{eqnarray}
F_{T,L}^{D(3),inel}(Q^2,x_P,\beta)=\frac{16e^2_f\alpha^5 N}{\pi} n_{eff}^2
\left(\frac{2a(x_P)}{\pi}\right)^3 x_P^{-1-2\Delta_P} \nonumber \\
\int_{c-i\infty}^{c+i\infty}\frac{d\gamma}{2\pi i}\left(\frac{r_0Q}{2}\right)^{\gamma}
\Omega(\gamma)H_{T,L}(\gamma) \beta^{-\Delta(\gamma)} \label{17a}
\end{eqnarray}
where $H_{T,L}$ are defined in (\ref{19}),
\begin{equation}
\Omega(\gamma)= V(\gamma) \frac{2}{\gamma(2-\gamma)^3} \frac{\Gamma^4(2-\gamma/2)\Gamma^2(1+\gamma/2)}{\Gamma(4-\gamma)\Gamma(2+\gamma)}
\label{18}
\end{equation}and
\begin{eqnarray}
V\left( \gamma \right) =\int_{0}^{1}\,_{2}F_{1}\left( 1\!-\!\gamma ,1\!-\!\gamma
;1;y^{2}\right) dy\ .  \label{25} 
\end{eqnarray}
($_{2}F_{1}$ is the hypergeometric function).
In the interesting 3-pomeron limit ($\beta<<1$) the path integral can be evaluated by the saddle point method with the result
\begin{eqnarray}
F_{T,L}^{D(3),inel}(Q^2,x_P,\beta)=G\ H_{T,L}(1)\frac{e^2_f\alpha^5 N^2\pi }{4}
\left(\frac{2a(x_P)}{\pi}\right)^3 x_P^{-1-2\Delta_P}
\frac{r_0Q}{2}  \nonumber \\
 \beta^{-\Delta_P} 
\left(\frac{2a(\beta)}{\pi}\right)^{\frac12} \exp\left(-\frac{a(\beta)}{2}\ln^2(r_0Q/2)\right)
\label{17}
\end{eqnarray}
where $G=.915...$ is Catalan's constant, $H_T(1)=9/2$, $H_L(1)=1$.

The important features of  Eq.(\ref{17}), pointed out in \cite{inelcomp} are

(a) An approximate factorization of the $x_P$ and $Q^2$ dependences.

(b) Important logarithmic corrections of the form $(\ln(1/x_P))^{-3}$ to the main power law factor $x_P^{-1-2\Delta_P}.$ These corrections lower the effective pomeron intercept for diffractive dissociation, in qualitative agreement with the data.

(c) There is a significant scaling violation, because $F_{T,L}^{D(3)}$ depends explicitly on $Q^2$. 
\vspace{.5cm}

(II) Quasi-elastic component.

It was discussed in \cite{elcomp}, where the  formulae 
for diffractive cross-sections in $\gamma^*$-dipole collisions were given. From these formulae one can derive  the following expression for the diffractive structure functions.

\begin{eqnarray}
F_T^{D(3),qel}(Q^2,x_P,\beta)\!\!\!\!\!\!&&=\frac{Q^{4}
N_{c}e_{f}^{2}}{2\pi^3\beta x_P}n_{eff}^2\int_{r_0}^{\infty} d^2b \int_{0}^{1}dz\left(z^2+(1-z)^{2}\right)z^{2}(1- z^{2}) \nonumber \\
&&\times \ \left| \int_{0}^{r_0}d\rho\ T\left(
b,\rho,r_0,\xi \right) K_{1}\left( \hat{Q}r\right) {J}_{1}\left( 
\hat{M}r\right)
\right| ^{2}  \label{32}
\end{eqnarray}
and

\begin{eqnarray}
F_L^{D(3),qel}(Q^2,x_P,\beta)  &&=\frac{Q^{4}
N_{c}e_{f}^{2}}{\pi^3\beta x_P}n_{eff}^2\int_{r_0}^{\infty} d^2b \int_{0}^{1}dz\ z^3 (1-z)^{3}  \nonumber \\
&&\times \ \left| \int_{0}^{r_0}d\rho\ T\left(
b,\rho,r_0,\xi \right)  K_{0}\left( \hat{Q}r\right) {J}_{0}\left( 
\hat{M}r\right)
\right| ^{2}  \label{32a}
\end{eqnarray}

where
\begin{equation}
 \hat{Q}^2 = z(1-z)Q^2, \;\; \hat{M}^2 = z(1-z)M^2  \label{33}
\end{equation}
and  $T(b,\rho,r_0,x_P)$ is the amplitude for elastic scattering of a dipole of diameter $\rho$ on a dipole of diameter $r_0$ at impact parameter $b$. 

In Ref.\cite{elcomp} this amplitude was approximated by its asymptotic form valid for large $b$ which reads
\begin{equation}
T(b,\rho,r_0,x_P)\approx \pi \alpha^2 \frac{\rho r_0}{b^2} \ln\left(\frac{b^2}{\rho r_0}\right)
x_P^{-\Delta_P}\left(\frac{2a(x_P)}{\pi}\right)^{\frac32} e^{\frac{a(x_P)}2}
\ln^2(\frac{b^2}{\rho r_0})    \label {33a} 
\end{equation}
and for that reason the integration over $b$ was performed from $r_0$ to $\infty$ (the meaning of the formula (\ref{33a}) for $b<r_0$ is rather doubtful).

 The main qualitative features of this quasi-elastic component, pointed out in \cite{elcomp}, are

(a) A similar $x_P$ dependence as the inelastic component, with important logarithmic corrections bringing down the pomeron intercept.

(b) As expected, the quasi-elastic component vanishes at $\beta =0$ and actually populates significantly only the region $\beta \geq .2$.

(c) The dependence on $\beta$ of the transversal and longitudinal structure functions is dramatically distinct. $F_T$ dominates in the region 
$\beta \leq .8$ whereas $F_L$ takes over at small $\beta$. The sum of the
two components, however, is almost constant in the range $.3 \leq \beta <1$.

\section{Predictions for hard diffraction}

The formulae presented in Section 4 were obtained in \cite{inelcomp,elcomp} by calculating first the cross-section of $\gamma^*$ on a single dipole of a fixed transverse diameter $r_0$ in the limit of very large impact parameter $b$. The obtained formulae were then extrapolated until $b_{min}=r_0$ and integrated from $b_{min}$ to $\infty$. Finally the result was multiplied by $n_{eff}^2$ to account for the number of the dipoles in the  target nucleon (determined from the fit of the formula for $F_2$ to the data).

These approximations allowed to perform explicit calculations and to discuss the general behaviour of diffractive structure functions \cite{inelcomp,elcomp}. They are, however, not valid in the important region where the impact parameter $b$ is of the order of the size of the original dipoles \cite {bia} and therefore the results given in the formulae of Section 3 cannot be treated
 as precise predictions of the QCD dipole picture (ie. of BFKL dynamics) for several reasons.

First, the asymptotic formula for large impact parameter ignores entirely the singularities of the dipole-dipole amplitudes, which become important when the impact parameter is of the order of the size of the colliding dipoles. This defect leads to a 
serious underestimation in the normalization of the calculated cross-sections\footnote{ It was recently shown in \cite{bia} that
this factor may  even well exceed 100.}. However, the conformal invariance of the BFKL dynamics \cite{li,Henri} insures that the general dependence on kinematic variables remains - to a good approximation- unaffected.

Second, the cross-section for scattering on a single dipole of the size $r_0$, even if multiplied by $n_{eff}^2$,  cannot be directly used for the estimation of the
cross-section on the nucleon target. The reason is twofold: (i) it is unlikely that all the primary dipoles in the nucleon are of the same size $r_0$ and thus the distribution of their sizes must be taken into account, (ii) The single-dipole cross-section 
ignores entirely the distribution of the transverse position of the primary dipoles in the nucleon, i.e. it ignores the effects of the nucleon form-factor. Although these effects are not present in forward scattering amplitudes (and therefore they do not 
influence the calculation of the total cross-section) they largely determine the momentum transfer dependence and thus reduce significantly the cross-section integrated over momentum transfer to the target nucleon.

To summarize, we note \cite {bia} two effects which were not included in the calculations given in \cite{inelcomp,elcomp} and which are expected to affect substantially the normalization of the obtained diffractive structure 
functions. In this situation  before
a  more precise calculation is available, we treat the normalization of the two components as free parameters, in order to phenomenologically evaluate the main conditions for  a test of the unified description of proton structure functions.  We thus  compare the experimental data to the formula
\begin{equation}
F_2^{D(3)} = N_{inel} F_2^{D(3),inel} + N_{qel} F_2^{D(3),qel}  \label{35}
\end{equation}
where  
$F_2^{D(3),inel}$ and  $F_2^{D(3),qel}$ are constructed from the formulae
(\ref{17}),  (\ref{32}) and (\ref{32a}) using $F_2=F_L+F_T$.

Since this procedure can at best be considered only as an exploratory search,
we did not try to perform a fit, but simply tried a few values of $N_{inel}$ and   $N_{qel}$ to see if one can obtain  a qualitative agreement of (\ref{35}) to the data. In Fig.4 the results of these calculations are shown for $N_{inel}=16$ and $N_{qel}=6$. One sees that a general description of the data is quite reasonable for $
x_P \leq .01$ except in the region of large $\beta$ where  the $Q^2$ dependence of the quasi-elastic component is not fully adequate.

We find this result rather satisfactory, given the present status of the theoretical calculations. Thus -although the final answer must wait till more precise QCD dipole calculations are available-
our tentative conclusion is that the existing data on rapidity-gap events do not rule out the BFKL dynamics as a correct description of the diffractive phenomena involving virtual photons. Indeed a decisive test will come along with more complete theoretical calculations e.g. \cite {bia}. 

\section{Conclusions and outlook}

In conclusion, we have shown that the BFKL dynamics, as represented by the QCD dipole picture, is is in qualitative agreement with the $3$-dimensional data on rapidity-gap events being observed at HERA. Further theoretical work is needed, however, to arrive at more precise conclusions. In particular, it is necessary:

(a) to evaluate the $\gamma^*$ cross-sections without the large-$b$ approximation used in \cite{inelcomp,elcomp}. The work on inelastic component was recently completed \cite{bia} and the quasi-elastic component will be available in the near future.

(b) Since the effects related to the nucleon form-factor are expected to influence significantly the results, a serious phenomenological discussion of the nucleon form-factor in the framework of the QCD dipole picture is required. More precise
 data on momentum transfer dependence of the diffractive structure 
functions would be of great help \footnote{ The form-factor effects being unimportant for forward scattering, the measurements of diffraction dissociation at zero momentum transfer would of 
course bring an important information to the problem we consider. At this point one may notice that also the measurements of the virtual photon shadowing in nuclei (which depends mainly on forward diffractive amplitudes \cite{Bialas1}) could provide another practical method to learn about the diffraction at zero momentum transfer.}.

We feel that this program is feasible and thus one may hope that a unified picture of the high-energy diffractive processes involving the virtual photons, based on BFKL dynamics, may indeed be constructed in the near future.

\eject

\newpage

{\bf FIGURE CAPTIONS}

\vspace{1cm}

{\bf Figure 1}
Fig.1a. Inelastic diffraction (component I)

\noindent Fig.1b. Quasi-elastic diffraction (component II)

\vspace{1cm}

{\bf Figure 2}
Comparison of the 4-parameter fit with the H1 data. The validity of the
prediction extends beyond the domain included in the fit. We note a
discrepancy at high $x$, high $Q^{2}$ due, in particular, to the absence of   the valence contribution
not considered  in the present model.

\vspace{1cm}

{\bf Figure 3}
Comparison of our prediction for the longitudinal structure function
$F_L$ and the H1 data. The prediction is somewhat lower than the
measurement, but more precise data are needed to make more precise tests.

\vspace{1cm}

{\bf Figure 4}
Prediction for the total  (longitudinal + transverse) diffractive structure function, see text. Dotted lines: the inelastic component I; Dashed lines the quasi elastic component II;  
Full line: the sum of both components (Note that at $\beta \approx 1,$ the inelastic component is almost 0 and the dashed line coincides with the full line and thus is not apparent on the plot).

\input epsf
\epsfysize=10.cm{\centerline{\epsfbox{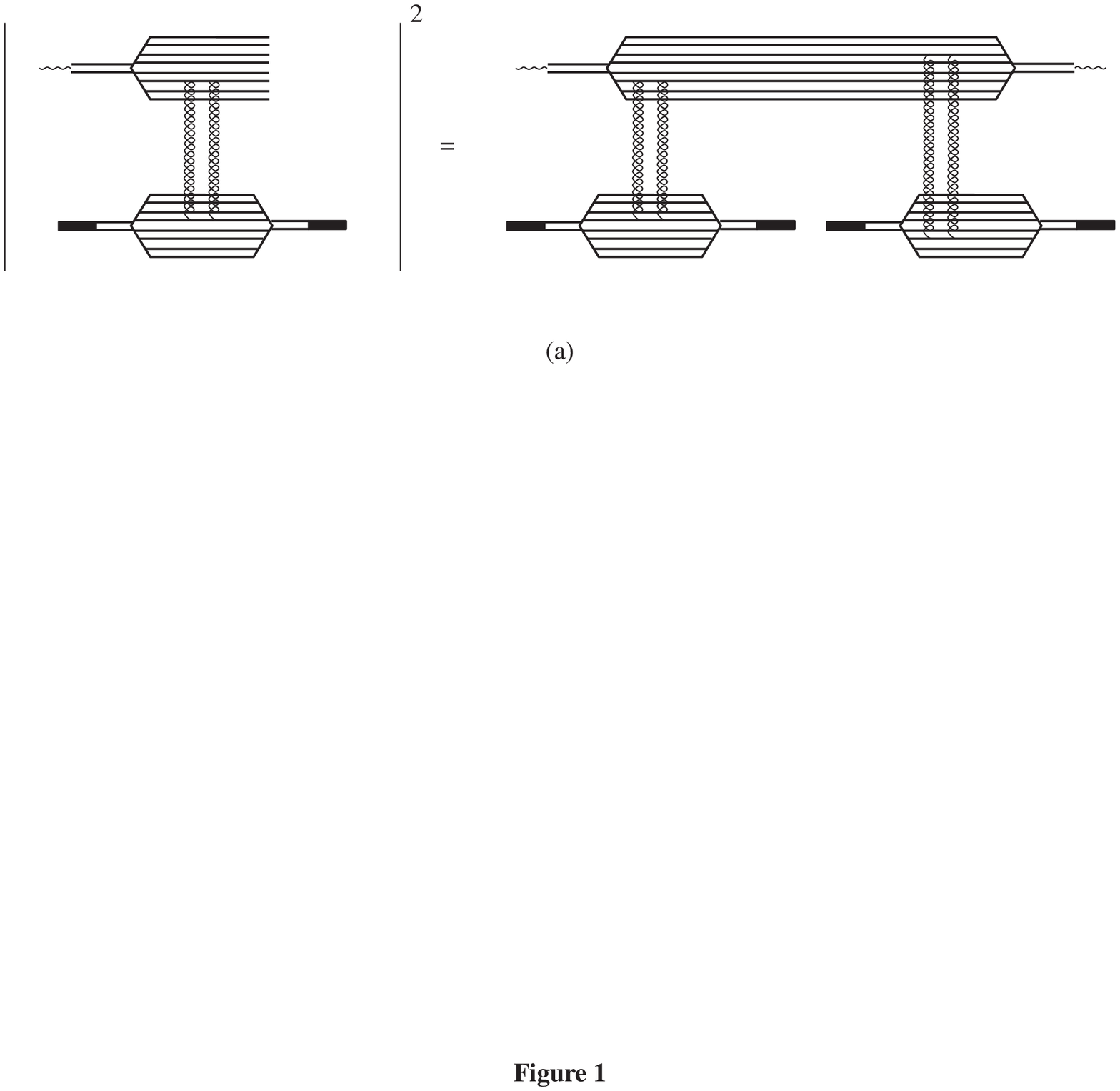}}}

\epsfysize=10.cm{\centerline{\epsfbox{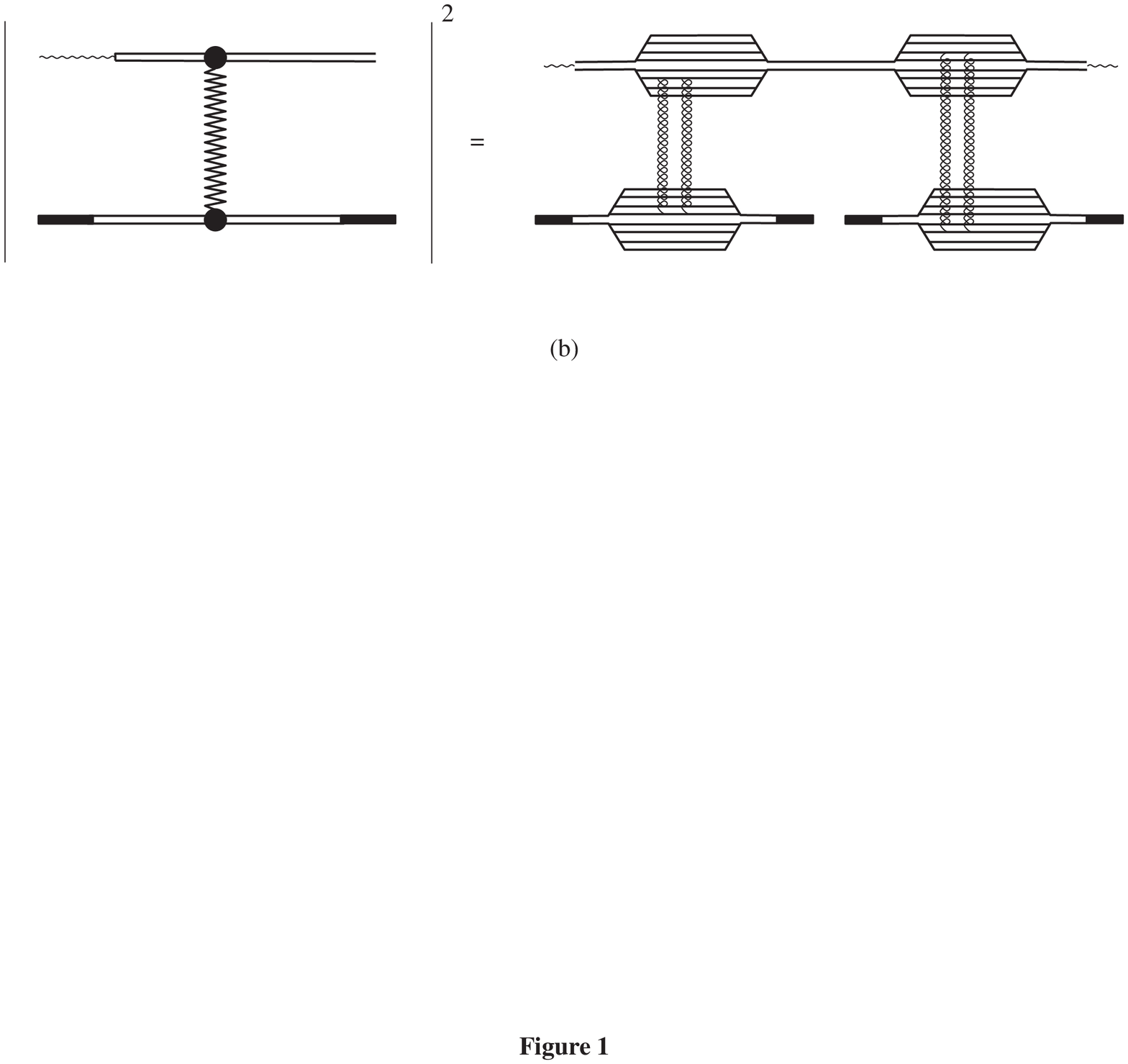}}}

\epsfysize=15.cm{\centerline{\epsfbox{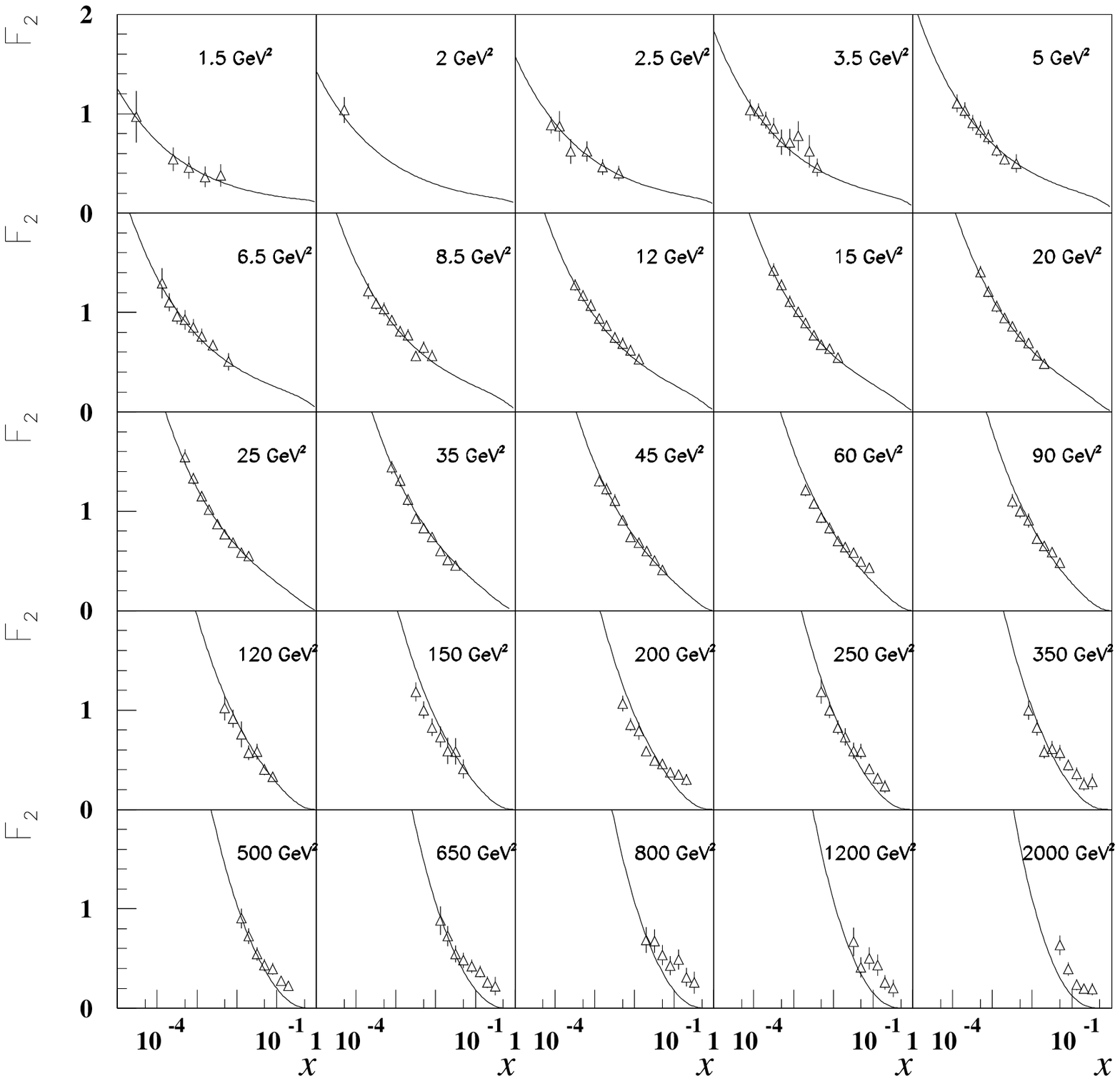}}}

\epsfxsize=15.cm{\centerline{\epsfbox{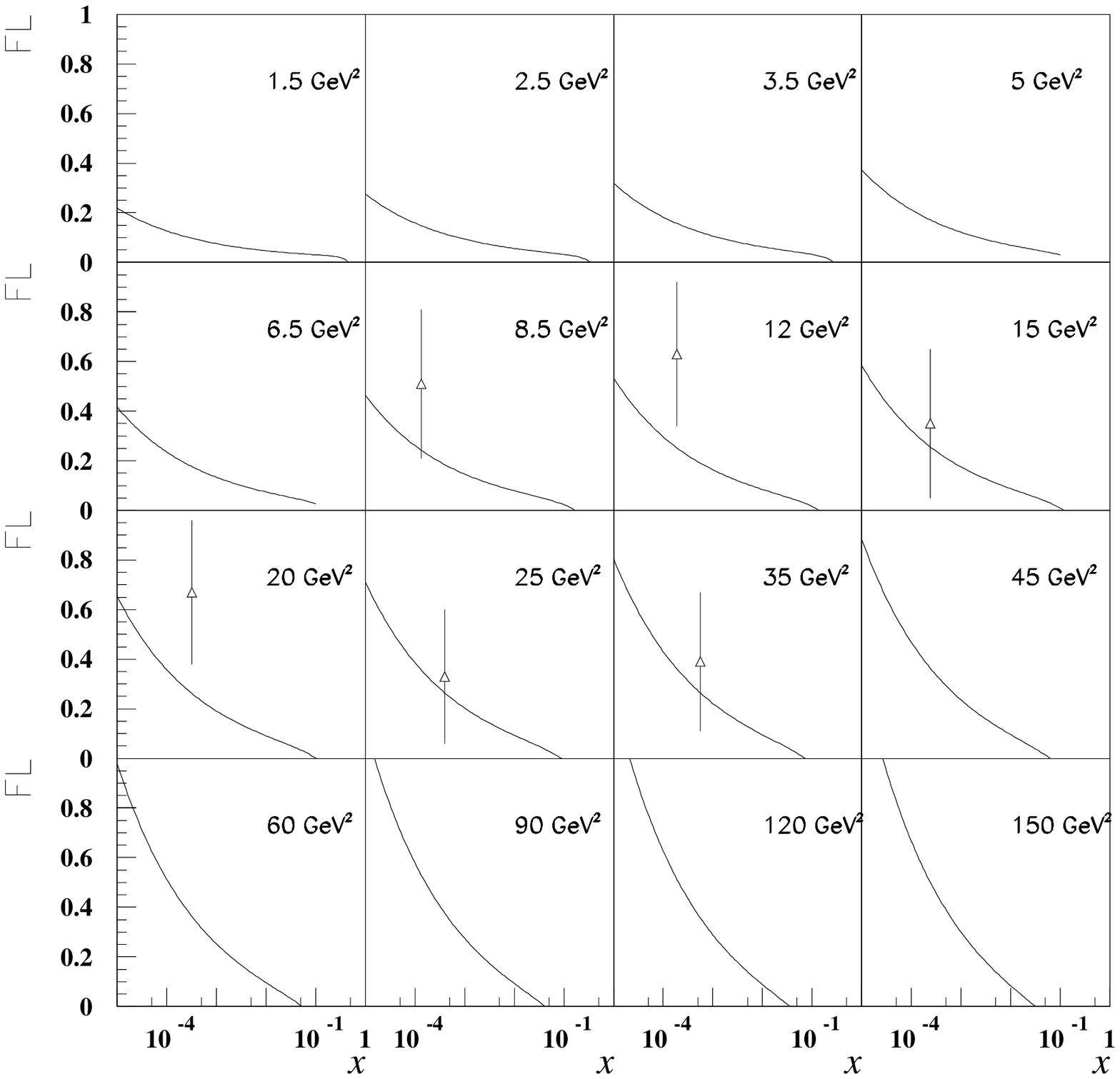}}}

\epsfxsize=17.cm{\centerline{\epsfbox{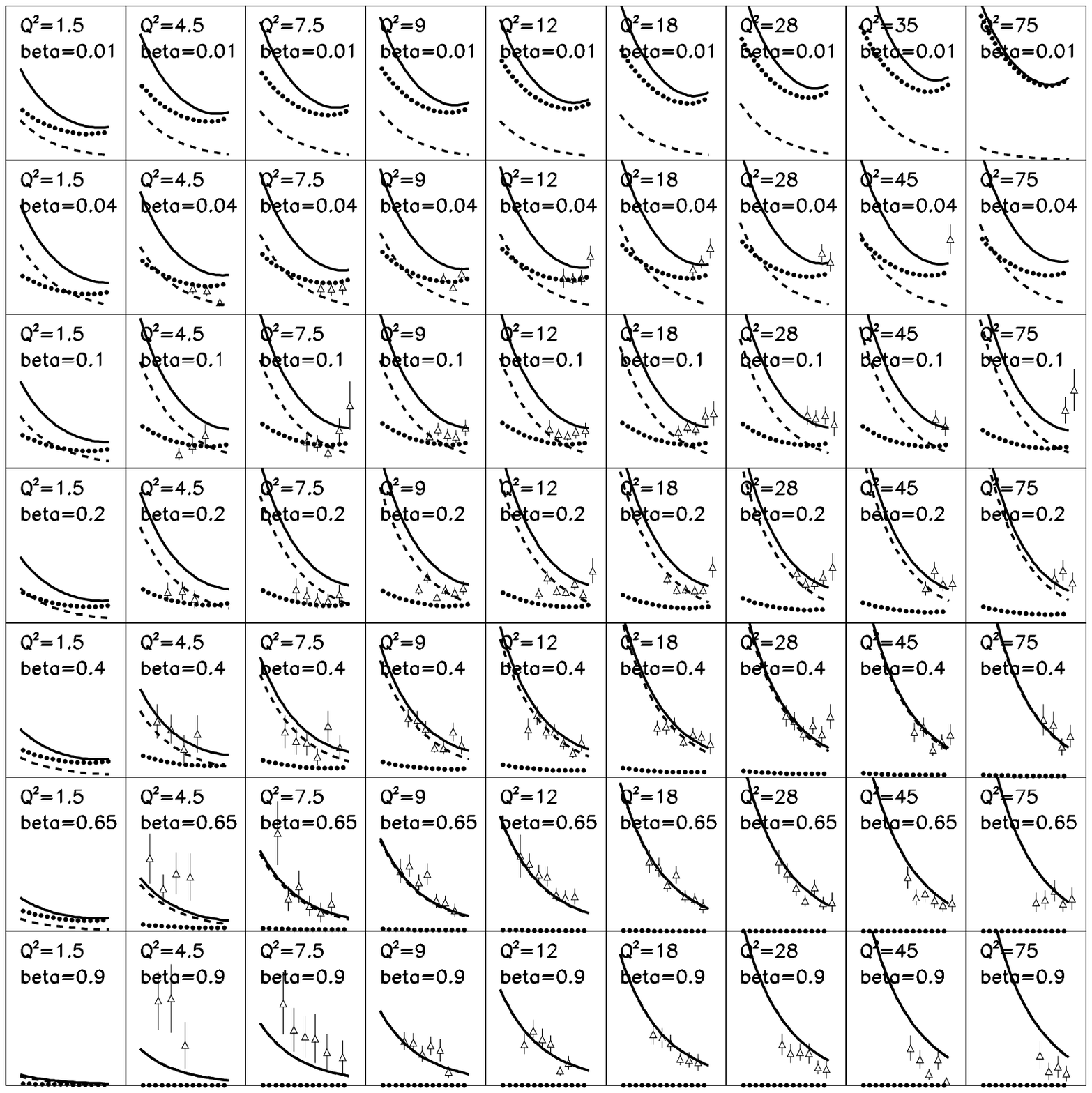}}}

\end{document}